\documentclass[prb]{revtex4}

\usepackage{graphicx}
\usepackage{dcolumn}
\usepackage{bm}
\usepackage{multirow}
\usepackage{xcolor}
\usepackage[normalem]{ulem}
\usepackage{amsmath}
\usepackage{amssymb}
\usepackage{layouts}
\usepackage{lipsum}
\usepackage{gensymb}
\usepackage[colorlinks=true]{hyperref}
\usepackage{booktabs}
\usepackage{threeparttable}

\begin{document}

\title[2D Excitonics with Atomically Thin Lateral Heterostructures]{2D Excitonics with Atomically Thin Lateral Heterostructures}

\author{S. Shradha$^1$, R. Rosati$^2$, H. Lamsaadi$^3$, J. Picker$^4$, I. Paradisanos$^5$, Md T. Hossain$^4$, L. Krelle$^1$, L. F. Oswald$^1$, N. Engel$^1$, D. I. Markina$^1$, K. Watanabe$^6$, T. Taniguchi$^7$, P. K. Sahoo$^8$, L. Lombez$^9$, X. Marie$^9$, P. Renucci$^9$, V. Paillard$^3$, J.-M. Poumirol$^3$, A. Turchanin$^4$,  E. Malic$^2$, B. Urbaszek$^1$}

\address{$^1$ Institute for Condensed Matter Physics, TU Darmstadt, Hochschulstraße 6-8, D-64289 Darmstadt, Germany}
\address{$^2$ Department of Physics, Philipps-Universität Marburg, Renthof 7, D-35032 Marburg, Germany}
\address{$^3$ CEMES-CNRS, Université de Toulouse, Toulouse, France}
\address{$^4$ Friedrich Schiller University Jena, Institute of Physical Chemistry, Lessingstr. 10, D-07743 Jena, Germany}
\address{$^5$ Institute of Electronic Structure and Laser, Foundation for Research and Technology-Hellas, Heraklion 70013, Greece}
\address{$^6$ Research Center for Functional Materials, National Institute for Materials Science, Tsukuba, Japan}
\address{$^7$ International Center for Materials Nanoarchitectonics, National Institute for Materials Science, Tsukuba, Japan}
\address{$^8$ Quantum Materials and Device Research Laboratory, Materials Science Centre, Indian Institute of Technology Kharagpur, Kharagpur, WB, India}
\address{$^9$ Université de Toulouse, INSA-CNRS-UPS, LPCNO, 135 Avenue Rangueil, 31077 Toulouse, France}

\vspace{10pt}
\begin{abstract}
Semiconducting transition metal dichalcogenides (TMDs), such as MoSe$_2$ and WSe$_2$, exhibit unique optical and electronic properties. Vertical stacking of layers of one or more TMDs, to create heterostructures, has expanded the fields of moiré physics and twistronics. Bottom-up fabrication techniques, such as chemical vapor deposition, have advanced the creation of heterostructures beyond what was possible with mechanical exfoliation and stacking. These techniques now enable the fabrication of \textit{lateral} heterostructures, where two or more monolayers are covalently bonded in the plane of their atoms. At their atomically sharp interfaces, lateral heterostructures exhibit additional phenomena, such as the formation of charge-transfer excitons, in which the electron and hole reside on opposite sides of the interface. Due to the energy landscape created by differences in the band structures of the constituent materials, unique effects such as unidirectional exciton transport and excitonic lensing can be observed in lateral heterostructures. This review outlines recent progress in exciton dynamics and spectroscopy of TMD-based lateral heterostructures and offers an outlook on future developments in excitonics in this promising system.   
\end{abstract}

\vspace{2pc}

\maketitle
%
%
%

\section{Introduction to lateral heterostructures in transition metal dichalcogenides}
Two-dimensional transition metal dichalcogenides (TMDs), denoted as $\text{MX}_2$ with M representing a transition metal atom and X a chalcogen atom, show very rich light-matter-interaction and exciton physics \cite{ROSATI2025312,barre2024engineering,choi2024emergence,Wang2018,Mueller18}. The presence of van der Waals (vdW) bonds between the layers of such materials has made it possible to not only exfoliate them down to monolayers but also to combine them into stacks known as vdW-heterostructures \cite{geim2013van}. 
Such vertical heterostructures may consist of one or more types of materials. A mismatch in the lattice constant and/or introducing a twist angle between the layers leads to the formation of superlattices with moiré potentials \cite{Seyler2019,Tran2019,Jin2019, Brem20c, Du2023}. Vertical heterostructures are hosts to phenomena such as interlayer excitons \cite{Rivera2015}, with the electron residing in one of the layers and the hole in another. This relies on the type of band alignment of the layers of the heterostructure. MoSe$_2$ and WSe$_2$ display a type-II (also called staggered) band alignment which is favourable for the formation of interlayer excitons with electrons in MoSe$_2$ and holes in WSe$_2$ \cite{Merkl19}. \\
\indent For excitonic transport in the monolayer plane, lateral heterostructures (LHs) are naturally more suitable. They consist of two or more monolayers that are covalently bonded to each other and show unique optical properties that we summarise in this work. The combination of materials that may be grown into LHs is dependent on the mismatch of the lattice constants of the constituent materials and the growth methods, which are discussed in the following sections. 

\begin{figure}[t]
\centering
\includegraphics[width=1\linewidth]{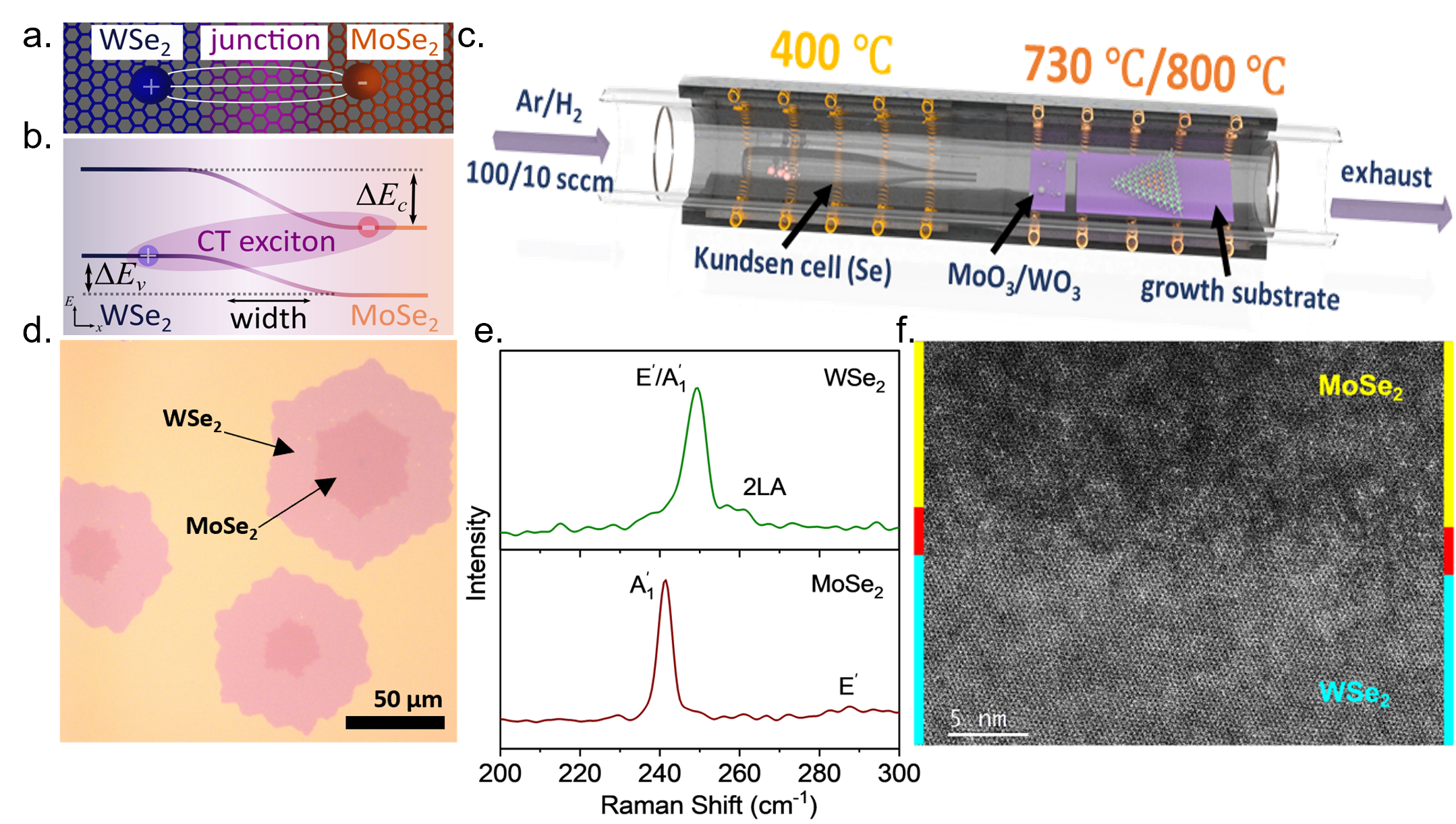} 
\caption{\textbf{ MoSe$_2$-WSe$_2$ lateral heterostructures} a. Sketch of the exemplary MoSe$_2$-WSe$_2$ lateral heterostructure and b. the corresponding in-plane variation of the single-particle energies, showing a type-II alignment favoring the emergence of charge transfer (CT) excitons. a., b. are adapted from \cite{Rosati2023}. c. Schematic of the CVD setup used for MoSe$_2$-WSe$_2$ growth on SiO$_2$/Si wafers with 300~nm of SiO$_2$. d. False-colored optical microscope images distinguishing MoSe$_2$ and WSe$_2$ domains. Reproduced with permission from \cite{Najafidehaghani2021}, Copyright 2021, Wiley-VCH GmbH. e. Raman spectrum of MoSe$_2$ (lower panel) showing the $\text{A}^{\prime}_1$ peak at 241~cm$^{-1}$ and Raman spectrum of WSe$_2$ (upper panel) with the $\text{E}^{\prime}$/$\text{A}^{\prime}_1$ peak at 250~cm$^{-1}$. f. HAADF-STEM image of the MoSe$_2$-WSe$_2$ interface, with MoSe$_2$ (yellow) and WSe$_2$ (cyan) domains as well as the boundary region (red) are highlighted. Reproduced with permission from  \cite{Beret2022}, Copyright 2022, Nature Publishing Group.}
\label{Fig-Growth}
\end{figure}

TMDs with low lattice-constant mismatch, such as MoSe$_2$ and WSe$_2$, can be grown into LHs having atomically sharp interfaces i.e. the changeover from one material to the other occurs within a few lattice constants \cite{ Beret2022, Najafidehaghani2021, Sahoo2019}. Unlike in vertical heterostructures where the layers are only connected via weak van der Waals bonds, the layers in LHs are covalently bonded. The presence of covalent bonds at atomically sharp interfaces, combined with a type-II band alignment, facilitates the transfer of excitons across the interface.  
A schematic representation of such a MoSe$_2$-WSe$_2$ LH is displayed in Figure~\ref{Fig-Growth}a. The ability to guide excitons in a given direction is important for excitonic devices, and will be summarized here for exciton transport across and along the LH interface. In standard TMD monolayers, excitons with large oscillator strengths are formed. However, control of these charge neutral particles is not straightforward as they do not directly respond to electric fields. However, this is different in the case of LHs, where the offset in exciton energies on the two sides of the junction spontaneously create an unidirectional exciton transport. 
Furthermore, the type-II band alignment in LHs such as MoSe$_2$-WSe$_2$ can, under certain conditions, lead to the formation of charge-transfer (CT) excitons that are localised at the interface. As depicted in Figure~\ref{Fig-Growth}b, these CT excitons consist of an electron on one side of the interface being bound to a hole residing in the material on the other side of the interface. 
There exist similarities between CT excitons and excitons in type-II quantum wells. For example in the GaAs/AlAs system, electrons are confined in the AlAs side and holes in the GaAs layer, resulting in excitons with a permanent electric dipole, as in the case of CT excitons \cite{Kesteren1990}. 
The formation of CT excitons is governed by modifications to the band offset, lateral junction width and the dielectric environment of the LH \cite{Rosati2023}.\\ 
\indent In light of the rapidly growing developments in the fields of material fabrication and precise optical characterization, the goal of this review is to present the recent advances in the understanding of optical properties of LHs with a focus on exciton dynamics at and across their interface. The following sections elaborate the current status of the growth techniques for the fabrication of LHs with focus on their excitonic properties.\\
\section{Growth of atomically thin lateral heterostructures}\label{sec:growth-Jena}
Initial attempts to engineer atomically thin heterojunctions were undertaken by top-down methods via mechanical exfoliation of individual TMD monolayers and their stacking in the van der Waals structures \cite{Lee2014}. However, these approaches are incapable of producing TMD LHs with atomically narrow boundaries; therefore, several bottom-up approaches were developed. In these methods, LHs are typically formed through the edge-epitaxy \cite{Duan2014,Huang2014,Li2015,Sahoo2018,Najafidehaghani2021,Zhang2018,Zhu2020}. 
Despite lattice mismatches between dissimilar TMD monolayers (e.g. 0.2~$ \% $ between MoSe$_2$ and WSe$_2$), this technique results in the nucleation and growth of a second TMD monolayer from the edges of a first one \cite{Duan2014,Huang2014,Li2015,Sahoo2018,Najafidehaghani2021,Zhang2018,Zhu2020,Gong2014,Turchanin2025}. In this way, atomically sharp in-plane junctions can be created that confine charge carriers and excitons directly along a 1D interface \cite{Li2015,Zhu2020,Beret2022,Rosati2023}. 
Several growth methods have been reported to this end, including multi-step chemical vapor deposition (CVD) \cite{Li2015,Zhang2018}, one-pot CVD \cite{Duan2014,Huang2014,Sahoo2018,Najafidehaghani2021,Gong2014} or pulsed-laser sulfur conversion \cite{Mahjouri-Samani2015}. In the following section, we provide a brief overview of these growth strategies, which are particularly relevant to exitonic phenomena. Note that a recent comprehensive review of the other growth methods for TMD LHs can be found in \cite{Liu2025}.
\subsection{Multi-step CVD and pulsed laser chalcogen conversion}
Multi-step CVD involves sequential growth of a second TMD domain on pre-grown domains of the first TMD. For example, Li et al. employed a multi-step CVD process to grow MoS$_2$ selectively along the edges of pre-deposited WSe$_2$ crystals, forming MoS$_2$-WSe$_2$ LHs \cite{Li2015}. In this technique, after the initial growth of the first TMD the CVD chamber is reloaded with materials for the growth of the second TMD or the initially grown TMD sample is transferred to a different CVD reactor for the subsequent growth. Beyond semiconductor-semiconductor junctions, metal-semiconductor LHs such as NbS$_2$-WS$_2$ have also been realized employing multi-step CVD \cite{Zhang2018}. These architectures, while primarily explored for electronic transport, hold promise for exciton-plasmon coupling at atomically defined boundaries, potentially enhancing light-matter interactions.\\
Alternative preparation strategies integrate top-down patterning with chemical conversion. For example, pulsed-laser-induced sulfurization has been used to convert selectively MoSe$_2$ domains into MoS$_2$, producing arrays of lateral junctions with typically 5~nm interface sharpness \cite{Mahjouri-Samani2015}. This method provides a platform for position-controlled exciton generation and routing, useful for excitonic circuits.

\subsection{One-pot CVD growth}
In contrast to multi-step approaches, one-pot CVD growth enables the sequential formation of TMD LHs within the same reaction chamber, offering a simple and scalable route as well as avoiding uncontrolled chemical modification of the 1D interface. Several studies have demonstrated the versatility of this approach by tailoring growth conditions. For instance, Duan et al. employed custom atmospheric pressure CVD systems with in situ switching of solid sources, using WS$_2$ and WSe$_2$ powders to grow WS$_2$-WSe$_2$, and MoO$_3$ with alternating S and Se to grow MoS$_2$-MoSe$_2$ LHs \cite{Duan2014}. 
A one-step heating growth method was also reported for WS$_2$-MoS$_2$ LHs at 650~℃ by combining MoO$_3$ and S powder with a tungsten-tellurium precursor; here, tellurium promoted tungsten melting \cite{Gong2014}. In another approach, MoSe$_2$-WSe$_2$ LHs were synthesized at 950~°C using mixed powders of MoSe$_2$ and WSe$_2$ exclusively under H$_2$ flow \cite{Huang2014}. Sahoo et al. further advanced the one-pot CVD approach by incorporating gas environment control and water vapor, enabling multi-domain growth of  MoSe$_2$-WSe$_2$ and MoS$_2$-WS$_2$ LHs \cite{Sahoo2018}.\\
Najafidehaghani et al. \cite{Najafidehaghani2021} demonstrated highly reproducible monolayer MoSe$_2$-WSe$_2$ LHs growth using a modified CVD approach with a Knudsen-type effusion cell for precise selenium delivery \cite{George2019}, while transition metal oxides served as Mo and W sources, as shown in Figure~\ref{Fig-Growth}c \cite{Turchanin2025}. The controlled supply of chalcogen precursors via the Knudsen cell ensured reproducible growth, offering a clear advantage over conventional open crucible methods. 
In this one-pot, two-step heating process, selective nucleation of MoSe$_2$ occurred at 730~°C due to the higher partial pressure of Mo species, followed by sequential lateral growth of WSe$_2$ at 800~°C. The resulting heterostructures, characterized by optical microscopy, Raman spectroscopy, and high-angle annular dark-field scanning transmission electron microscopy (HAADF-STEM), Figure~\ref{Fig-Growth}d-f, exhibited atomically sharp interfaces \cite{Najafidehaghani2021}. Such well-defined interfaces are essential for minimizing disorder and enabling clean observation of 1D interfacial excitons, such as CT excitons \cite{Rosati2023}. 
Liquid precursors have also been explored as a source of transition metals for the one-pot growth of WS$_2$-MoS$_2$ LH for electrical transport studies \cite{Zhu2020}, and MoSe$_2$-WSe$_2$ LH for photodetector applications \cite{Zhao2024}, although their excitonic properties remain largely unexplored.\\
\begin{figure}[b]
\centering
\includegraphics[width = 1\linewidth]{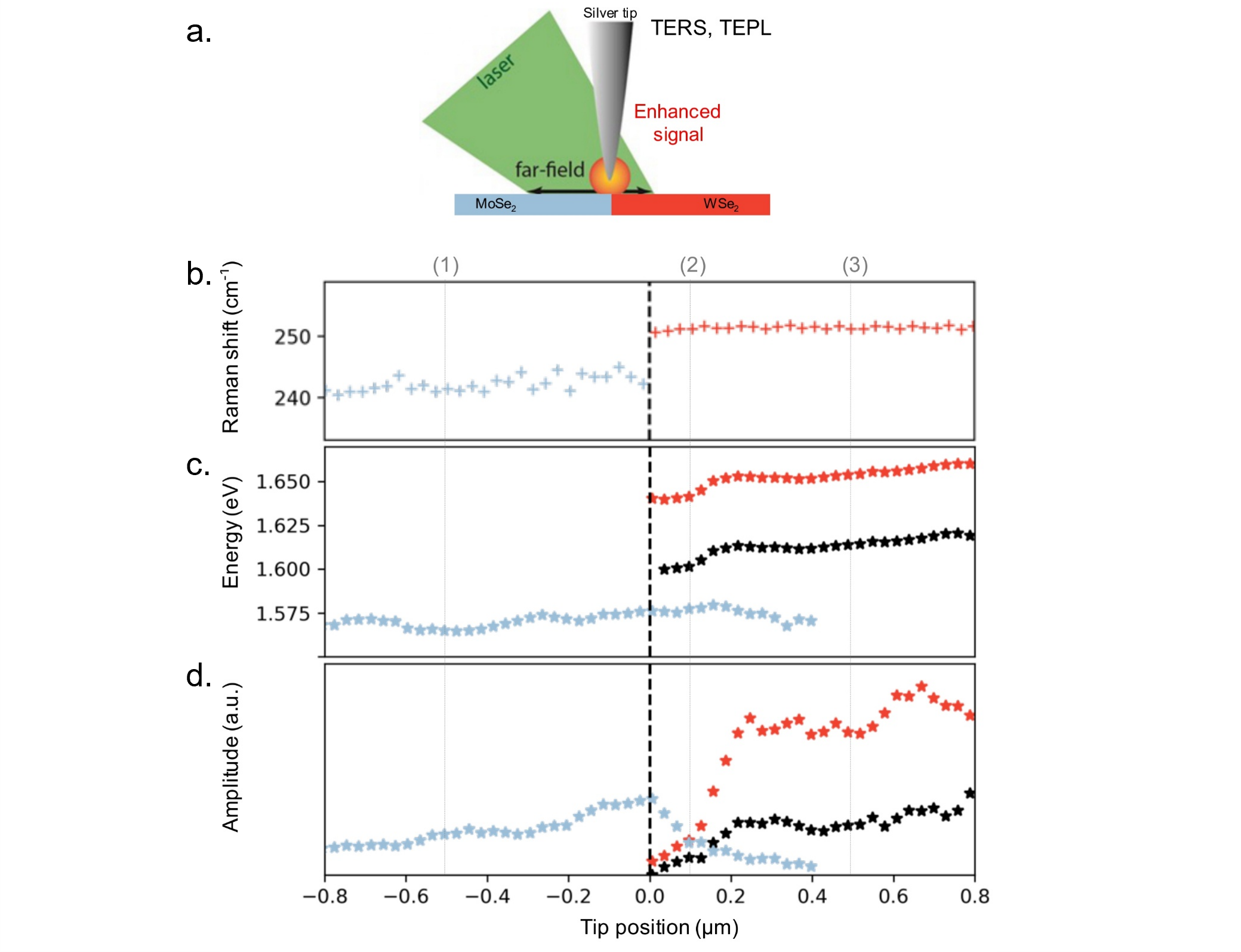} 
\caption{\textbf{Tip-enhanced spectroscopy of the MoSe$_2$-WSe$_2$ interface at T = 300~K}. a. Schematic of TERS and TEPL setup. b. Raman shift measured by TERS shows the $\text{A}^{\prime}_1(\Gamma)$ phonon line. The position of the interface is indicated with the black dashed line. c. PL emission energy obtained from Lorentzian fits of individual peaks. On the left of the interface the MoSe$_2$ neutral exciton (blue stars) contribution is visible while on the right of the interface the WSe$_2$ neutral (red stars) and dark (black stars) excitonic contributions are seen. d.  Amplitude of the individual peaks obtained from TEPL shown in c. Figures reproduced from \cite{Beret2022}, Copyright 2022, Nature Publishing Group.}
\label{Fig-TERS-TEPL}
\end{figure}
In summary, the described approaches highlight the versatility of CVD and related methods in producing 2D-TMD LHs with. While multi-step CVD and pulsed-laser conversion provide flexibility in the material design and spatial control, one-pot CVD enables scalable growth of lateral junctions with structurally seamless, yet atomically sharp and chemically well-defined interfaces. 
By avoiding exposure to the ambient environment, one-pot CVD offers a straightforward and reproducible route to high-quality 2D-TMD LHs. Conducting systematic exciton spectroscopy on these structures is essential to link advances in synthesis with excitonic device applications, paving the way for exploring rich 2D excitonic physics in next-generation optoelectronic devices.\\
\section{Excitons in lateral heterostructures}\label{sec:exciton-TUD}
Several optical spectroscopy techniques can be used to study the role of excitons in optical absorption and emission in  TMD monolayers and heterostructures \cite{shree2021guide}. Photoluminescence (PL), Raman and differential reflection (DR) spectroscopy are used as key tools to probe the optical properties of LHs. 
These techniques shed light on the exciton dynamics at the interface of LHs.
Below, the key findings are presented.\\
\subsection{Photoluminescence and Raman Spectroscopy}
PL spectroscopy is used to probe excitonic properties at and across the interface of LHs, as has been shown by several studies in the field \cite{Beret2022,Kundu2024}. Due to the narrow width (2-3~nm) of the interface in certain LHs, tip-enhanced spectrosocopy techniques characterized by high spatial resolution, have been employed to characterize LHs \cite{Beret2022}. A schematic of the setup for room temperature tip-enhanced PL (TEPL) and tip-enhanced Raman spectroscopy (TERS) of a MoSe$_2$-WSe$_2$ LH is presented in Figure~\ref{Fig-TERS-TEPL}a. Here the excitation laser is focused on a silver tip with a diameter of about 40~nm, which roughly corresponds to the spatial extent of the excited region. 
PL and Raman spectral scans obtained in this manner are displayed in Figure~\ref{Fig-TERS-TEPL}b and c. 
In Figure~\ref{Fig-TERS-TEPL}b, a sharp shift of the position of the $\text{A}^{\prime}_1{(\Gamma)}$ Raman mode is seen. Thus the position of the interface (represented by the dotted line) can be determined by TERS. It has also been demonstrated that second-harmonic generation (SHG) imaging can be used to reveal atomically sharp interfaces in MoSe$_2$–WSe$_2$ LHs, showing a ~23$\%$ enhancement in SHG intensity at the junction due to coherent interference between the second-order nonlinear responses of the adjoining monolayers \cite{sousa2021revealing}. 

Figure~\ref{Fig-TERS-TEPL}c tracks the position of the PL peaks obtained from TEPL. 
In the absence of an efficient transport across the interface, an optical excitation on the WSe$_2$ (MoSe$_2$) side should lead to a PL being clearly dominated by the emission from WSe$_2$ (MoSe$_2$) excitons. However, unlike in the case of the Raman modes, where a sharp changeover can be seen at the interface, in PL, contribution from the $\text{X}_{0}^{\text{MoSe}_2}$ exciton emission is observed upto 400~nm into the WSe$_2$ region of the LH. 
This behaviour is also depicted clearly by plotting the PL amplitude of the individual Lorentzian peaks obtained by fitting the PL spectra (Figure~\ref{Fig-TERS-TEPL}d). This indicates that excitons formed in WSe$_2$ transfer preferentially through the interface into the MoSe$_2$ region where they recombine radiatively. As further discussed in Section~\ref{sec: exciton-prop-across}, such uni-directional exciton transport is driven by the offset in the exciton energies at the two sides of the junction, which can be furthermore in competition with the trapping into CT excitons \cite{rosati2025}. This transfer causes near the interface the quenching of WSe$_2$-related PL and the enhancement of the MoSe$_2$-related PL, well above the one observed in MoSe$_2$ far from the junction as the excitons are pumped from the high-gap TMD toward the low-gap TMD.
Further investigation of LHs has been carried out at 4K, as at low temperatures the contributions of the different excitonic species (bright, dark and trions) are well separated.\\ 
\begin{figure}
\centering
\includegraphics[width=1\linewidth]{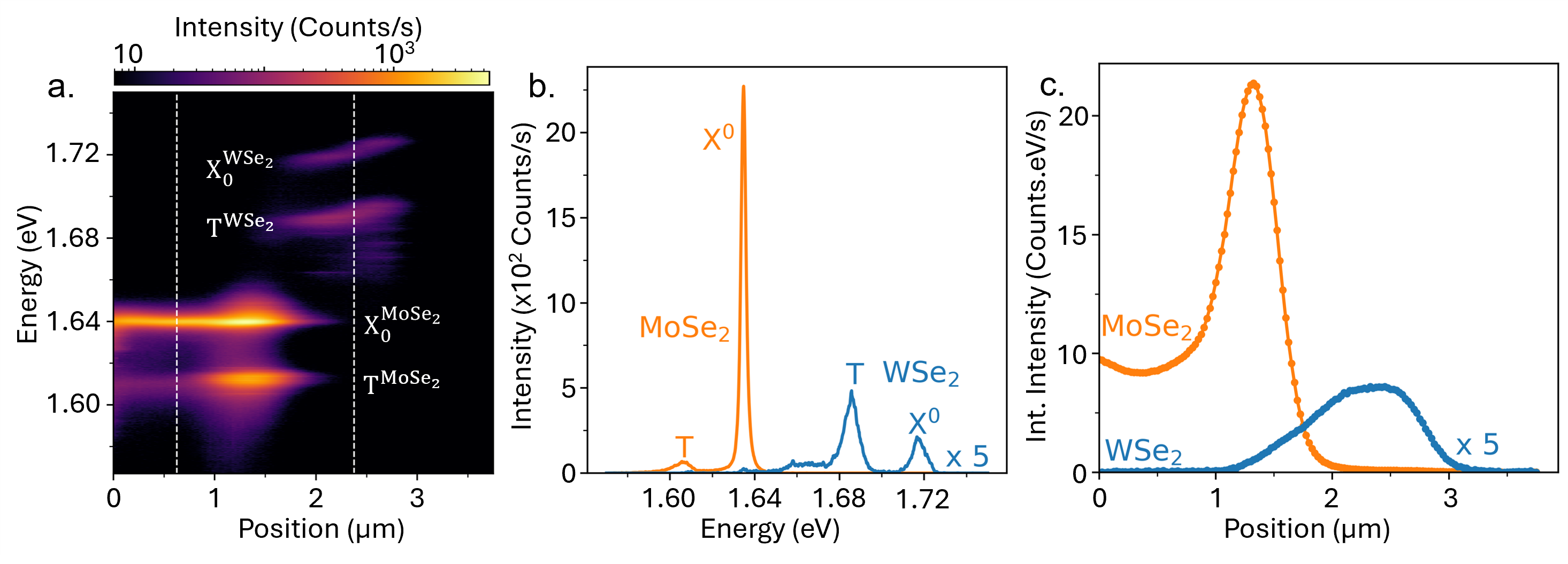} 
\caption{\textbf{Photoluminescence from a MoSe$_2$-WSe$_2$ lateral heterostructure encapsulated in hBN at T = 4.7~K} a. PL line scan across the interface of the LH. The neutral excitons ($\text{X}_0^{\text{MoSe}_2}$, $\text{X}_0^{\text{WSe}_2}$)  and, the trions ($\text{T}^{\text{MoSe}_2}$, $\text{T}^{\text{WSe}_2}$) from the two materials of the LH are marked. The white dashed lines indicate the positions whose individual spectra are plotted in panel b. b. Spectra obtained in monolayer regions away from the interface showing MoSe$_2$ (orange) and WSe$_2$ (blue). MoSe$_2$ PL is dominated by emission from the neutral exciton $\text{X}_0$ at 1.638~eV and a weaker contribution from the trion (\text{T}) at 1.605~eV. The PL intensity from WSe$_2$ is characteristically lower than that of MoSe$_2$. In this case, the trion at 1.685~eV has a higher intensity compared to the neutral exciton at 1.717~eV. c. The cumulative (T + $\text{X}_0$ )  integrated PL from MoSe$_2$ (orange) and WSe$_2$ (blue) are plotted as a function of position in the line scan. Upon approaching the interface of the two materials, the intensity from MoSe$_2$ increases relative to that at positions away from the interface. These measurements are performed with a 633~nm He:Ne laser and a diffraction limited excitation spot.}
\label{Fig-PL}
\end{figure}
Figure~\ref{Fig-PL}a shows a typical heatmap of a PL line scan of  a MoSe$_2$-WSe$_2$ LH at a temperature of 4.7~K with a diffraction limited excitation spot. 
In the MoSe$_2$ and WSe$_2$ regions, two main emission lines are usually observed. The respective energies of the emission lines in MoSe$_2$ remain rather constant throughout the scan. However, one can observe a redshift in the WSe$_2$ peaks close to the interface between the materials, indicating the presence of a residual local strain. Line cuts from this PL scan are shown in Figure~\ref{Fig-PL}b. 
In the MoSe$_2$ monolayer region of the LH, the trion (T) and neutral exciton ($\text{X}_0$) are observed. In the WSe$_2$ region of the LH, the $\text{X}_0$ is observed. About 32~meV below it, a broad and brighter peak is found. This is attributed to the trion \cite{Wang2015}. In this particular sample, not all WSe$_2$ spectral features \cite{Yang2022} are resolved. Due to the spin-dark nature of the energetically lowest-lying states in WSe$_2$, the cumulative intensity of the emission from WSe$_2$ is lower relative to that of MoSe$_2$ at T = 4.7~K. 
This is visible in Figure~\ref{Fig-PL}c, where the blue curve indicates the cumulative WSe$_2$ intensity and the orange one represents MoSe$_2$. From this plot, it is also evident that there is a sharp increase in the intensity of the MoSe$_2$ emission near the interface to about four times the intensity in monolayer regions away from the interface. This increase stems from the type-II band alignment between MoSe$_2$ (lower bandgap) and WSe$_2$ (higher bandgap), the high quality and the atomically sharp nature of the interface. Upon excitation of WSe$_2$, excitons that form in the WSe$_2$ layer find it energetically favourable to move across the interface before recombining as MoSe$_2$ excitons \cite{Lamsaadi2023}.

\indent In summary, the interface position is determined by techniques such as Raman spectroscopy and in PL spectroscopy at 4 K and at room temperature, the transfer of excitons from WSe$_2$ to MoSe$_2$ is observed, as will be discussed in the next sections in detail. This is possible due to a LH interface with a minimal number of defects that permits exciton transport.
\subsection{Differential Reflection Spectroscopy}
\begin{figure}[h]
\centering
\includegraphics[width = 1\linewidth]{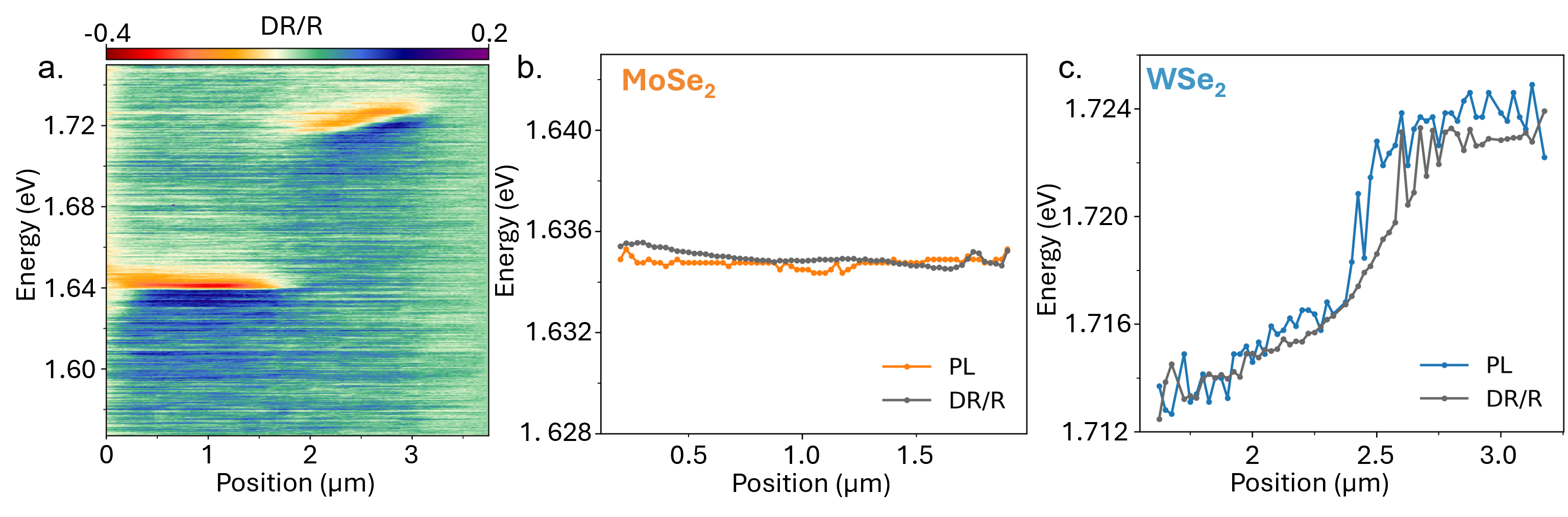} 
\caption{\textbf{Differential white light reflectivity DR/R at lateral junction at  T=4.1~K}. a. Line scan of differential white light reflectivity DR/R measured across the interface of a $\text{MoSe}_2-\text{WSe}_2$ LH (same sample as Figure~\ref{Fig-PL}). From position 0~µm to 2~µm, a resonance at 1.635~eV from MoSe$_2$ is observed, and from 2~µm onwards the resonance corresponding to WSe$_2$ appears, indicating that the strong oscillator strengths of these resonances are well maintained in the LH. b. DR resonance (black) and $X_0^{\text{MoSe}_2}$ (orange). Energies obtained from both of these measurements overlap well and appear fairly constant both in regions away from and close to the interface. c. DR resonance (black) and emission energy of $X_0^{\text{WSe}_2}$ (blue). Both of these show a position-dependent energy shift of about 8~meV. }
\label{Fig-DR}
\end{figure}
DR spectroscopy  has been used as an efficient technique to study resonances with high oscillator strength in 2D-TMDs \cite{shree2021guide,raja2019dielectric}. In nominally undoped samples, mainly the neutral excitons are observed with DR spectroscopy. Figure~\ref{Fig-DR} shows the results of a DR line scan across the same MoSe$_2$-WSe$_2$ LH interface as in Figure~\ref{Fig-PL}. In Figure~\ref{Fig-DR}a, a heatmap of the line scan shows the presence of two distinct resonances in the structure. 
The lower energy resonance at 1.635~eV is $X_0^{\text{MoSe}_2}$ while the higher energy resonance corresponds to $X_0^{\text{WSe}_2}$. As observed in the case of the PL emission from WSe$_2$, the position of the resonance has a gradual redshift close to the interface of the LH. 
This is better visualised in Figure~\ref{Fig-DR}c. Here, the PL emission energy (in blue) as well as the DR resonance (in black) of WSe$_2$ are plotted as a function of the position of the line scan. PL and DR spectral features exhibit a gradual shift of about 8~meV lower in the vicinity of the interface. The PL (in orange) and DR resonance (in black) of the MoSe$_2$ region of the LH are shown in Figure~\ref{Fig-DR}b. Both PL and DR resonance energies are observed to have a constant value of 1.635~eV at positions close to as well as away from the interface.\\ 
The selective redshift observed in WSe$_2$ emission upon approaching the interface, is attributed to strain at the interface. This occurs due to the inherent nature of the growth of the LHs which is discussed in detail in the previous section. MoSe$_2$ with a lattice constant of 3.288~\AA~\cite{James1963} is grown first. In the second step, WSe$_2$, whose lattice constant is 3.282~\AA~\cite{Schutte1987}, is grown. The lattice mismatch of these materials is $\sim$ 0.2~\%. This is sufficient for the WSe$_2$ lattice to experience a tensile strain in the vicinity of the interface. 
As the  growth of the WSe$_2$ continues away from the interface, the monolayer can relax back to its unstrained lattice parameters, thus shifting the resonance and corresponding emission energy to higher values. The excellent agreement of the transition energies measured in DR and PL serve as an indicator of high-quality growth with low defect concentrations and a sharp interface. 

\subsection{Charge-transfer excitons}
\begin{figure}
\centering
\includegraphics[width=1\linewidth]{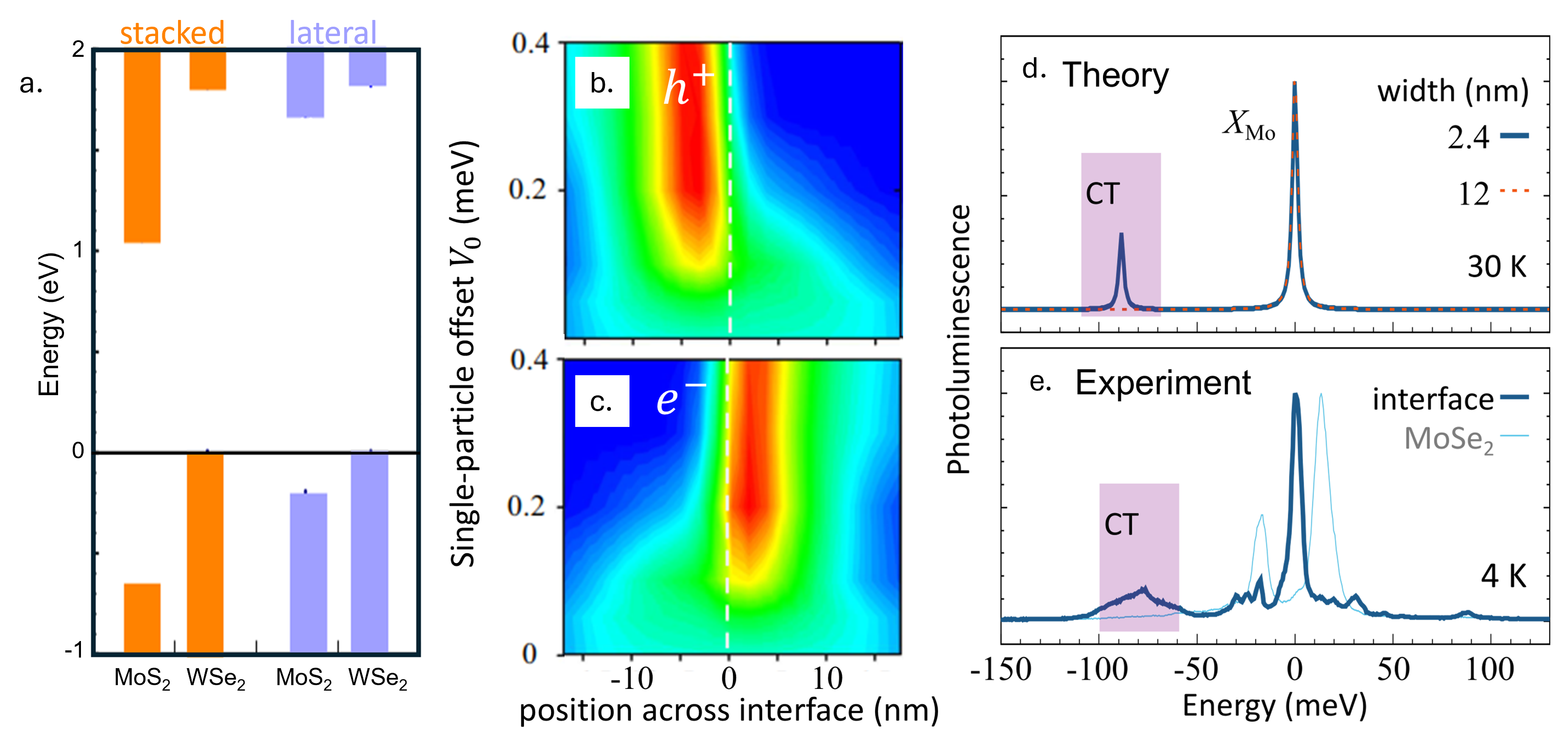} 
\caption{\textcolor{black}{\textbf{Charge transfer excitons in lateral heterostructures}  a. Band alignment in vertical and lateral heterostructures, showing a drastic reduction of the band offset in the latter. b.-c. Offset-driven localization of  electrons and holes at the opposite sides of the interface thanks to the appearance of CT excitons. d. Theoretically predicted and e. experimentally measured cryogenic  photoluminescence after an optical excitation at the interface. The appearance of a new CT exciton peak X$_{\text{CT}}$ located approximately 90 meV below the monolayer X$_{\text{Mo}}$ exciton is observed - in excellent theory-experiment agreement. Figures~a., b.-c. and d.-e., are adapted respectively from \cite{Guo16}, \cite{Lau18}, and \cite{Rosati2023}.}}
\label{Fig-CT}
\end{figure}

Lateral heterostructures show spatially dependent single-particle energy bands and exciton energies, as they are composed of different materials on the two sides of the interface \cite{Guo16,Kang2015,Lau18,Bellus18,Rosati2023,Yuan23} as shown in Figure~\ref{Fig-Growth}a and b. 
This gives rise to peculiar spectra in line scans across the interface presented in Sec. \ref{sec:exciton-TUD}, and also drives a remarkable carrier transport across the interface as will be discussed in Sec. \ref{sec:transport} \cite{Beret2022,Lamsaadi2023,Lamsaadi2025,Bellus18,
Shimasaki22,Kundu2024,Zhong24,Kundu2024_arxiv}.
The exemplary MoSe$_2$-WSe$_2$ lateral heterostructure shows a type-II band alignment, facilitating the formation of CT excitons, with electrons and holes spatially separated in the energetically favorable MoSe$_2$ and WSe$_2$ side, respectively, cf. Figure~\ref{Fig-Growth}b.
This is in analogy to the vertically stacked van der Waals heterostructures, where interlayer excitons formed by spatially separated carriers are studied in detail \cite{Rivera2015,Ciarrocchi19,shree2021guide,Schmitt22}. In contrast, CT excitons have been demonstrated only recently in lateral MoSe$_2$-WSe$_2$ heterostructures \cite{Rosati2023,vandoolaeghe25} (cf. Figure~\ref{Fig-CT}e) and in WS$_{1.16}$Se$_{0.84}$-WSe$_2$ heterostructures \cite{Yuan23}. 
CT excitons can form the energetically lowest energy state at the junction and therefore play a crucial role in the transfer and conversion of excitons between the WSe$_2$ and MoSe$_2$ sides.
To better understand CT excitons, we now discuss their special characteristics
 and how their optical response can be controlled by interface and dielectric engineering of lateral heterostructures.
The single-particle band alignment shows important quantitative differences between vertical and lateral heterostructures. The exemplary case of  MoS$_2$-WSe$_2$ heterostructures shows a type-II alignment in both cases \cite{Guo16}, however with the minimum offset reduced by almost a factor of 4 for LHs (220 meV compared to 810 meV for the corresponding vertical heterostructures), cf. Figure~\ref{Fig-CT}a.  
This difference stems from the role of the charge-neutrality level, to which the band energies are pinned only in the case of LHs \cite{Guo16}. Nevertheless, already at  this relatively small offset, CT electron-hole pairs represent the energetically lowest states \cite{Lau18, Rosati2023}.\\ 
The corresponding CT exciton wave function shows the localization of electrons and holes at opposite sides of the interface, cf. Figure~\ref{Fig-CT}b and c.  Such excitons show static electric dipoles corresponding to electron-hole separations of several nanometers. This is one order of magnitude larger than the dipole moment of interlayer excitons in vertical heterostructures, where the spatial dipole is geometrically restricted by the layer separation \cite{Lau18, Rosati2023}.  
Note that CT exciton transition energy is below  the monolayer intralayer excitons only for band offsets $V_0=(\Delta E_c + \Delta E_v)/2\gtrsim$100 meV. Below this value, the smaller bandgap of the unbound CT continuum compared to the MoSe$_2$ bandgap is not able to compensate the fact that CT excitons have smaller binding energies than  monolayer excitons due to the large electron-hole separation \cite{Rosati2023}.
This threshold of approximately 100 meV is typically smaller than  realistic band offsets in lateral  heterostructures, cf. Figure~\ref{Fig-CT}a. The visibility of CT excitons is, however, also crucially affected by other factors, such as the interface width and the dielectric environment,  as we discuss below.\\
Excitons can recombine into photons only when electron and hole wavefunctions overlap in real space. As a consequence, the oscillator strength of CT excitons is reduced for increasing dipoles (electron to hole separation). This depends on the width of the interface, as induced by spontaneous alloying during the growth process. When the interface width becomes larger than the Bohr radius of the CT exiton, the oscillator strength is suppressed \cite{Rosati2023}.
Modern CVD growth techniques limit the spontaneous alloying to small interface widths of only 2-3 nm \cite{Najafidehaghani2021,Rosati2023,Beret2022,Li2015,Xie18,Ichinose22}. This has allowed the observation of CT excitons in hBN-encapsulated WSe$_2$-MoSe$_2$ lateral heterostructures, with a new low-energy peak appearing roughly 90 meV below the monolayer  $X_{\text{Mo}}$ peak - in excellent agreement between theoretical predictions and experimental observations, cf. Figures~\ref{Fig-CT}d and e. Such a peak is  absent when exciting away from the interface (cf. thin line in Figure~\ref{Fig-CT}e) or in the case of broader interface widths (orange dashed line in Figure~\ref{Fig-CT}d) reflecting an increased spatial dipole and hence suppressed oscillator strength. Different samples or even positions of the same structure could hence result  in a negligible CT exciton photoluminescence \cite{Rosati2023}. 
Interestingly, while free-standing mono- or bilayers show a rich photoluminescence spectrum \cite{Kumar24}, in lateral WSe$_2$-MoSe$_2$ heterostructures, the CT exciton peak is absent in free-standing samples \cite{Rosati2023}. 
This occurs because exciton energies depend on their binding energy, which becomes larger for a reduced screening of the Coulomb interaction \cite{Wang2018,ROSATI2025312}. The increase is, however, larger for intralayer excitons than for the charge-transfer ones, as the latter present overall much smaller binding energies due to the spatial separation between electron and hole. This leads  to a decrease of the energy separation between CT and intralayer excitons upon reduction of the dielectric screening (with values of approximately 88, 35 and 6 meV for hBN-encapsulated, supported on SiO$_2$ and free-standing WSe$_2$-MoSe$_2$ lateral heterostructures, respectively \cite{Rosati2023}). A smaller energy separation results in a weaker CT exciton occupation and thus to its reduced visibility in PL spectra.\\
\subsection{Exciton propagation along the interface }\label{sec:transport}
\begin{figure}
\centering
\includegraphics[width = 1\linewidth]{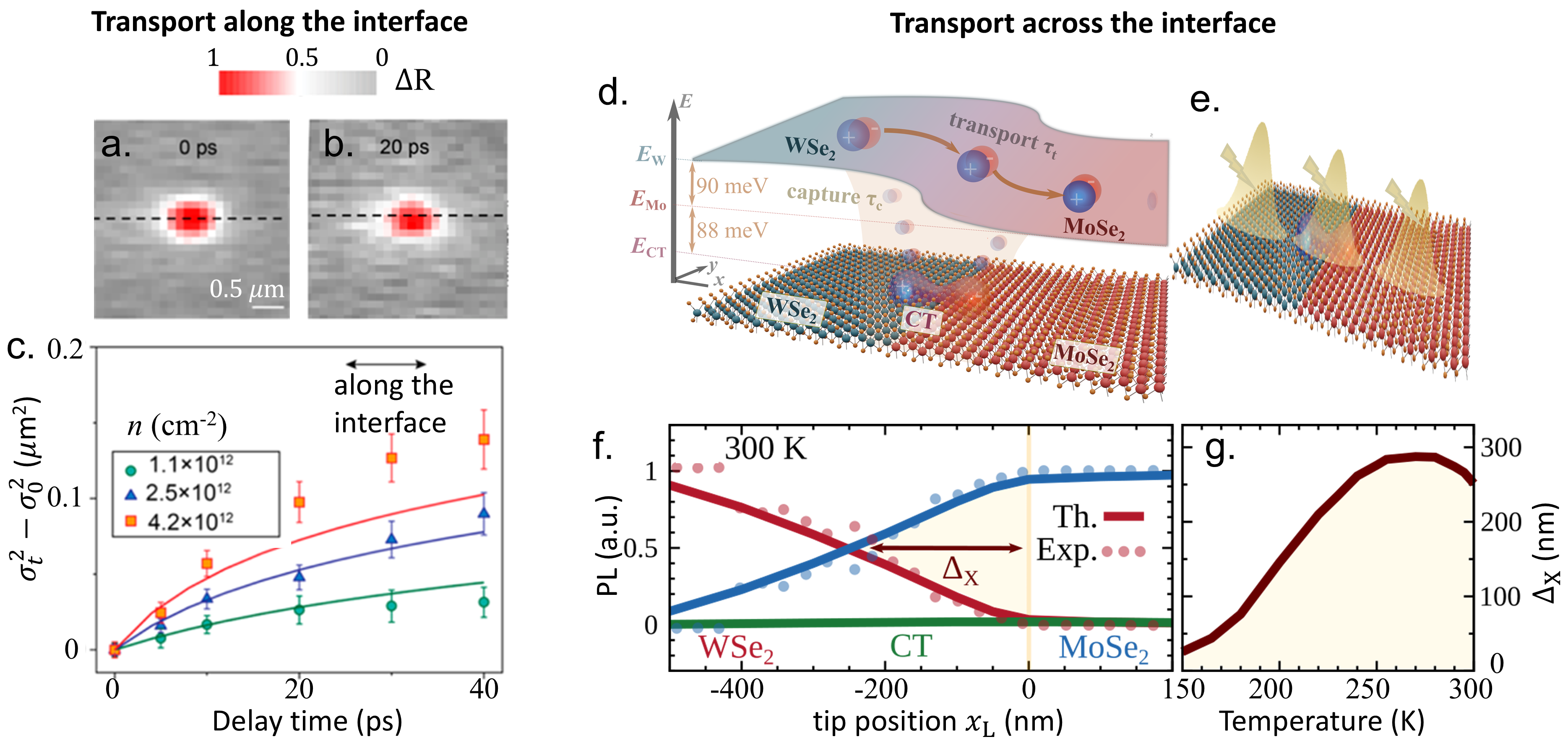} 
\caption{\textbf{Exciton transport in lateral heterostructures}. Spatially-resolved differential reflectivity a. 0 ps and b. 20 ps after a localized excitation at the interface of a WS$_{1.16}$Se$_{0.84}$-WSe$_2$ lateral heterostructure, showing the formation of an inhomogeneous exciton distribution elongated along the interface (dashed line). c. Density-dependent evolution of the mean-squared displacement along the interface, revealing a non-linear exciton diffusion. d. Sketch of the  offset-driven exciton drift across the interface and trapping into CT excitons directly at the interface. e. Sketch of a near-field raster scanning experiment. f. Intensity of the photoluminescence emitted by MoSe$_2$, WSe$_2$ and CT-excitons (blue, red and green) as a function of the near-field tip position (with the interface at $\text{x}_\text{L}=0$ as denoted by the vertical line). The interplay of the  offset-driven unidirectional exciton propagation and the capture into CT excitons breaks the symmetry, resulting in a spatial offset $\Delta_{\text{X}}$ from the interface, at which the emission from WSe$_2$ and MoSe$_2$ excitons is equal - in excellent agreement between theory and experiment. g. Such a spatial offset sensitively depends on temperature and is strongly suppressed at small temperatures, reflecting an efficient trapping of WSe$_2$ excitons, before they could reach the MoSe$_2$ side. Figures~a.-c. and d.-g. are adapted respectively from \cite{Yuan23} and \cite{rosati2025}.}
\label{Fig-transport}
\end{figure}
\begin{figure}[h]
    \centering
    \includegraphics[width=0.48\linewidth]{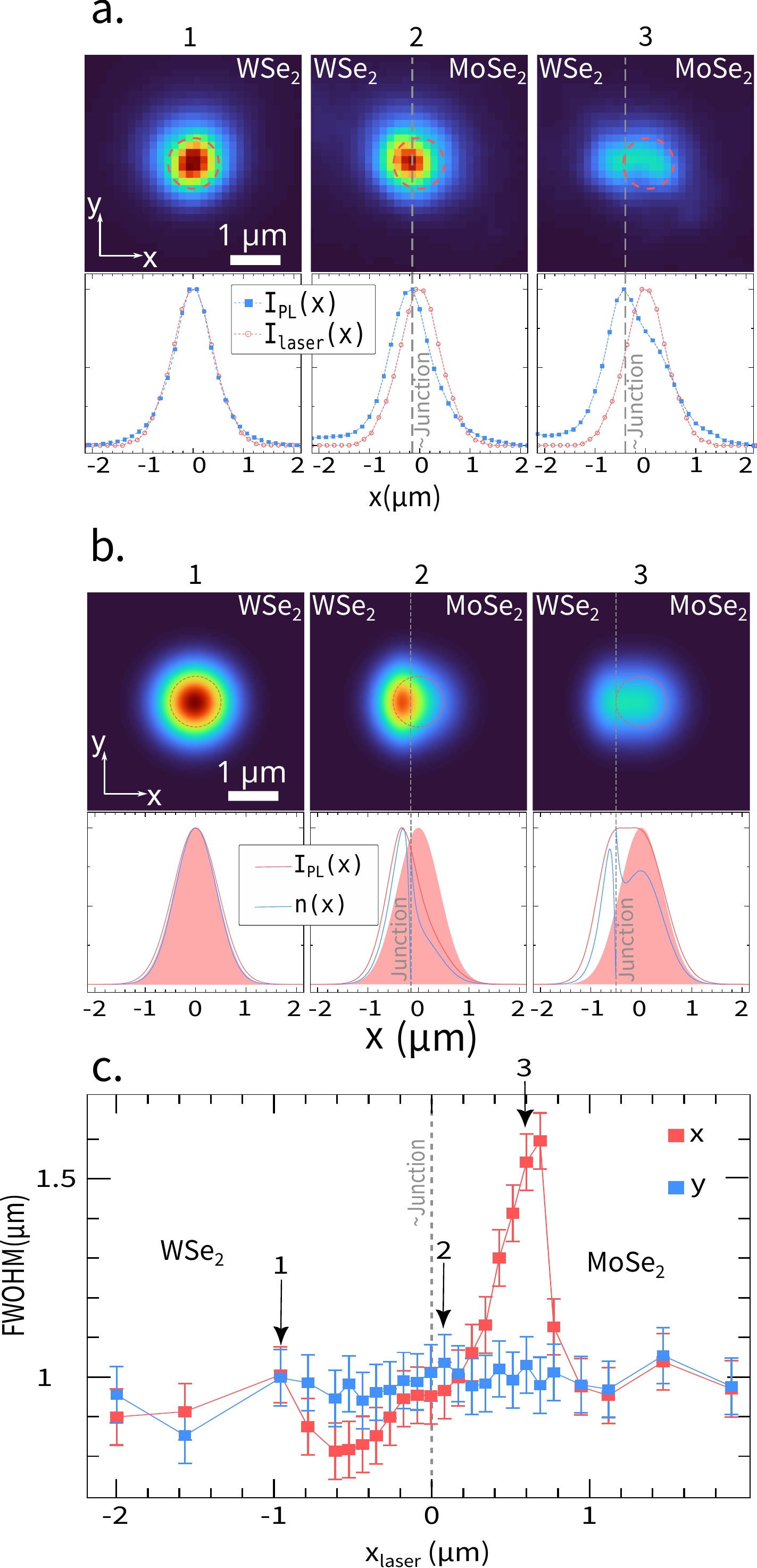}
    \caption{\textbf{Far-field optical imaging of WSe$_2$–MoSe$_2$ LHs at room temperature.} a. Top: µPL images recorded far from the junction (1) inside WSe$_2$ and near it at (2) and (3). Color maps are individually normalized. 
    The dashed red circle marks the laser spot, and the dashed gray line indicates the estimated junction position. Bottom: $x$-profiles from the PL maps. Blue: integrated PL intensity, $I_{PL}(x,x_{laser})=\int dy I_{PL}(x,y,x_{laser})$; red: laser profile, $I_{laser}(x)=\int dy I_{laser}(x,y)$. b. Same as a., but obtained by solving the 2D diffusion equation. c. Full Width at Outer Half Maximum (FWOHM) of $I_{PL}(x,x_{laser})$ (red) and $I_{PL}(y,x_{laser})$ (blue) versus laser position. FWOHM is the distance between the outermost points where the intensity halves. Adapted from \cite{Lamsaadi2025}.}
    \label{ExcitonLensing1}
\end{figure}
Lateral heterostructures host intralayer excitons with different energies on the two sides of the interface and they can exhibit energetically lowest CT excitons at the interface. The interplay between these three exciton species leads to a highly remarkable transport behaviour both along and across the interface. 
Optically excited monolayer excitons close to the interface are trapped into the energetically favorable CT exciton states. This results in unusually high CT-exciton densities that can be 1-2 orders of magnitude larger than monolayer exciton densities a few tens of picoseconds after the excitation \cite{rosati2025}. This high density together with the large spatial dipole moment can explain the highly non-linear exciton transport measured along the interface of a WSe$_2$-WS$_{1.16}$Se$_{0.84}$ lateral heterostructure \cite{Yuan23}.  
After an initial optical excitation at the interface, pump-probe measurements reveal anisotropic exciton profiles that are elongated along the interface, cf. Figures~\ref{Fig-transport}a and b. 
The evolution of the mean square displacement of the spatial profile  deviates from the linear increase expected for conventional diffusion \cite{Kulig18,Perea19,Rosati20} and it furthermore shows a considerable speed-up of the diffusion with increasing pump power, cf. Figure~\ref{Fig-transport}c  - analogous to the case of vertical heterostructures \cite{Yuan20,Sun2022,Erkensten22,Malic23,ROSATI2025312, Tagarelli23}. 
The effective diffusion coefficient increases up to few tens of cm$^2$/s for excitation densities of 4.5$\times 10^{12}$cm$^{-2}$, which is up to two orders of magnitude larger than in the case of  exciting the same sample away from the interface.
This increase in the effective diffusion coefficient is attributed to the strong dipole-dipole repulsion between CT excitons\,\cite{Yuan23}.
This finding suggests that lateral heterojunctions act as one-dimensional ``highways'' for the transport of excitons along the interface.\\
\subsection{Exciton propagation across the interface}
\label{sec: exciton-prop-across}
Directional exciton transport could be technologically relevant, but it is hindered by the neutral nature of excitons, which inhibits the use of electric fields. In regular mono- or bilayers, directional exciton funneling is possible in presence of strain fields \cite{Rosati21e,Cordovilla18,Harats20,Moon20,Lee22,Gelly22}, inhomogeneous gates in vertical heterostructures \cite{Unuchek18}, in anisotropic dielectric environments \cite{wang2025sub10nmnanochannels} or by integrating monolayers with complex photonic structures \cite{Jiang2025}. 
In contrast, lateral heterostructures naturally support exciton drift due to  the presence of a spatial energetic offset  \cite{Beret23,Bellus18,Lamsaadi2025}, cf. Figure~\ref{Fig-transport}d. 
This technologically promising behaviour can be revealed by different experimental techniques such as those discussed in Section~\ref{sec:exciton-TUD}. In near-field raster scanning microscopy, a nanometer exciton distribution is excited in different position across the interface by means of a metallic tip, cf. Figure~\ref{Fig-transport}e. The resulting space- and time-integrated PL spectrum reveals the exciton propagation by comparing the intensity of the PL peaks emitted by WSe$_2$ and MoSe$_2$ excitons as a function of the tip position, Figure~\ref{Fig-transport}f. Considering the full spatiotemporal exciton dynamics, it is found that interestingly the MoSe$_2$ emission dominates even for laser positions clearly on the WSe$_2$ side (up to a spatial offset $\Delta_X\approx$250 nm away from the interface) -  in excellent agreement between theory and experiment \cite{rosati2025,Lamsaadi2023}, cf. Figure~\ref{Fig-transport}f. Such a spatial offset $\Delta_X$, which would be zero in the absence of unidirectional transport across the interface, is five times larger than the initial localization of the laser pulse. \\
\indent In the vicinity of the interface on an LH, excitons experience two competing processes, unidirectional transport and trapping in CT exciton states.
Such trapping takes place locally \cite{Glanemann05,Reiter07,Rosati17} at the interface, where CT excitons are localized. The capture process becomes more efficient at low temperatures due to a suppressed escape of trapped excitons into the continuum of intralayer exciton states \cite{Reiter07}. As a consequence of the increased trapping, excitonic drift becomes weaker at smaller temperatures, in direct contrast to the behaviour in conventional semiconductors \cite{rosati2025}. 
This results in an overall decrease of the spatial offset away from the interface, which becomes smaller by one order of magnitude when going from 300 K to 150 K. Interestingly, a small decrease of the spatial offset is observed for temperatures above approximately 280~K (cf. Figure~\ref{Fig-transport}g). The non-monotonous behaviour of the spatial offset as a function of temperature can be traced back to the interplay of exciton trapping and drift, which show an opposite  temperature dependence.\\
\textcolor{black}{Theoretical predictions reveal how the unidirectional exciton drift could also be observed for far-field excitations in (i) space-resolved PL measuring the exciton accumulation at the MoSe$_2$ side \cite{Lamsaadi2023, rosati2025}, and in (ii) time-resolved PL measuring the increased height of the MoSe$_2$ PL peak relative to the WSe$_2$ peak \cite{rosati2025}. Furthermore, exciton transport can be controlled by dielectric and interface engineering. 
Here, smaller dielectric constants and smaller band offsets reduce the impact of CT excitons, suppressing capture processes at the interface and boosting the unidirectional transport across the interface \cite{rosati2025}.}

Using TEPL and a modified exciton transfer model, one can show a discontinuity of the exciton density distribution on each side of the interface. 
This introduces the concept of exciton Kapitza resistance, in analogy to the interfacial thermal resistance \cite{RevModPhys.41.48}. By comparing different heterostructures with or without top hexagonal boron nitride (hBN) layer, the transport properties can be controlled, over distances far greater than the junction width, by the exciton density through near-field engineering and/or laser power density, as summarized in \cite{Lamsaadi2023}. Surprisingly, a larger spatial offset is measured in uncapped lateral MoSe$_2$-WSe$_2$ heterostructure compared to the fully encapsulated sample \cite{Lamsaadi2023}, which needs to be further investigated with regards to dielectric screening \cite{rosati2025} and exciton-exciton scattering \cite{Zipfel2020}.\\

\subsection{Exciton focusing and lensing}
As described above, at room temperature, exciton diffusion near a lateral heterojunction has very specific characteristics : (i) it occurs from high-gap to low-gap and (ii) it occurs non-isotropically. This suggests that depending on the junction geometry, one can profoundly modify the trajectory of excitons in a controllable way and achieve complex in-plane manipulation of exciton flux and distribution.
Exciton distributions and flux orientations across LHs have been
examined using room temperature µPL imaging of the emission patterns \cite{Lamsaadi2025}. As shown in Figure~\ref{ExcitonLensing1}a (image 1), far from the junction, exciton diffusion appears isotropic, with symmetric PL pattern centered at the excitation laser spot. In contrast, when the laser excites both WSe$_2$ and MoSe$_2$ at the interface, the PL distribution becomes highly asymmetric. The position of the maximum PL emission shifts away from the laser center, and the spatial extent of the PL spot becomes strongly dependent on the laser position. Specifically, the PL spot becomes narrower than the laser spot when the latter is centered near the interface (cf. Figure~\ref{ExcitonLensing1}a, image 2), while it broadens when the laser spot mostly illuminates the WSe$_2$ side (cf. Figure~\ref{ExcitonLensing1}a, image 3). Note that WSe$_2$ exhibits higher PL intensity at room-temperature \cite{Wang2015}.  The full set of PL pattern modifications near the junction, can be reproduced by theoretical modeling. This is achieved by solving a classical 2D diffusion equation that incorporates bandgap-induced exciton drift at the interface (cf. Figure~\ref{ExcitonLensing1}b) \cite{Lamsaadi2025}. \\
To investigate the directional influence of LHs on exciton dynamics and fluxes, µPL imaging can be performed along a line perpendicular to the interface. The spatial broadening is quantified using the full width at outer half maximum (FWOHM), defined as the separation between the most distant points where the PL intensity reaches 50\% of its maximum. 
Figure~\ref{ExcitonLensing1}c displays the FWOHM of the µPL spot along the $x$-axis (red curve) and the $y$-axis (blue curve) as a function of the laser spot position. The PL intensity profiles along the $x$-axis, extracted from the µPL images and defined as $I_{PL}(x,x_s) = \int dy I_{PL}(x,y,x_s)$, are recorded as a function of the laser excitation position, $x_{laser}$. 
Near the junction, these profiles clearly split into two distinct peaks, with the emission maximum noticeably shifted away from the center of the excitation spot  (cf. Figure~\ref{ExcitonLensing1}c, red squares). In contrast, the corresponding $y$-axis profiles, defined as $I_{PL}(y,x_s) = \int dx I_{PL}(x,y,x_s)$, consistently exhibit a single peak centered at the laser position (cf. Figure~\ref{ExcitonLensing1}c, blue squares). 
This directional asymmetry indicates that an atomically sharp lateral heterointerface strongly modifies the exciton distribution perpendicular to the interface, while having negligible effect along the parallel direction. Notably, two distinct diffusion regimes emerge in the vicinity of the junction:
\begin{itemize}
    \item A forced diffusion regime is identified, in which the FWOHM along the $x$-axis becomes smaller than that along the $y$-axis (i.e., the red curve lies below the blue one). The FWOHM decreases to values as low as $800$~nm, indicating that the µPL spot along the $x$-direction becomes narrower than the excitation laser spot itself. This counterintuitive behavior points to the presence of strongly anisotropic exciton diffusion, wherein excitons are driven toward regions of higher density. Such behavior is indicative of exciton condensation and is associated with a negative effective diffusion length.
    \item An enhanced diffusion regime arises, where the FWOHM along the $x$-axis increases by up to 50\% relative to that along the $y$-axis, reflecting a pronounced extension of exciton propagation across the junction and a strong enhancement of the effective diffusion length.
\end{itemize}
\begin{figure}[!t]
    \centering
    \includegraphics[width=1\linewidth]{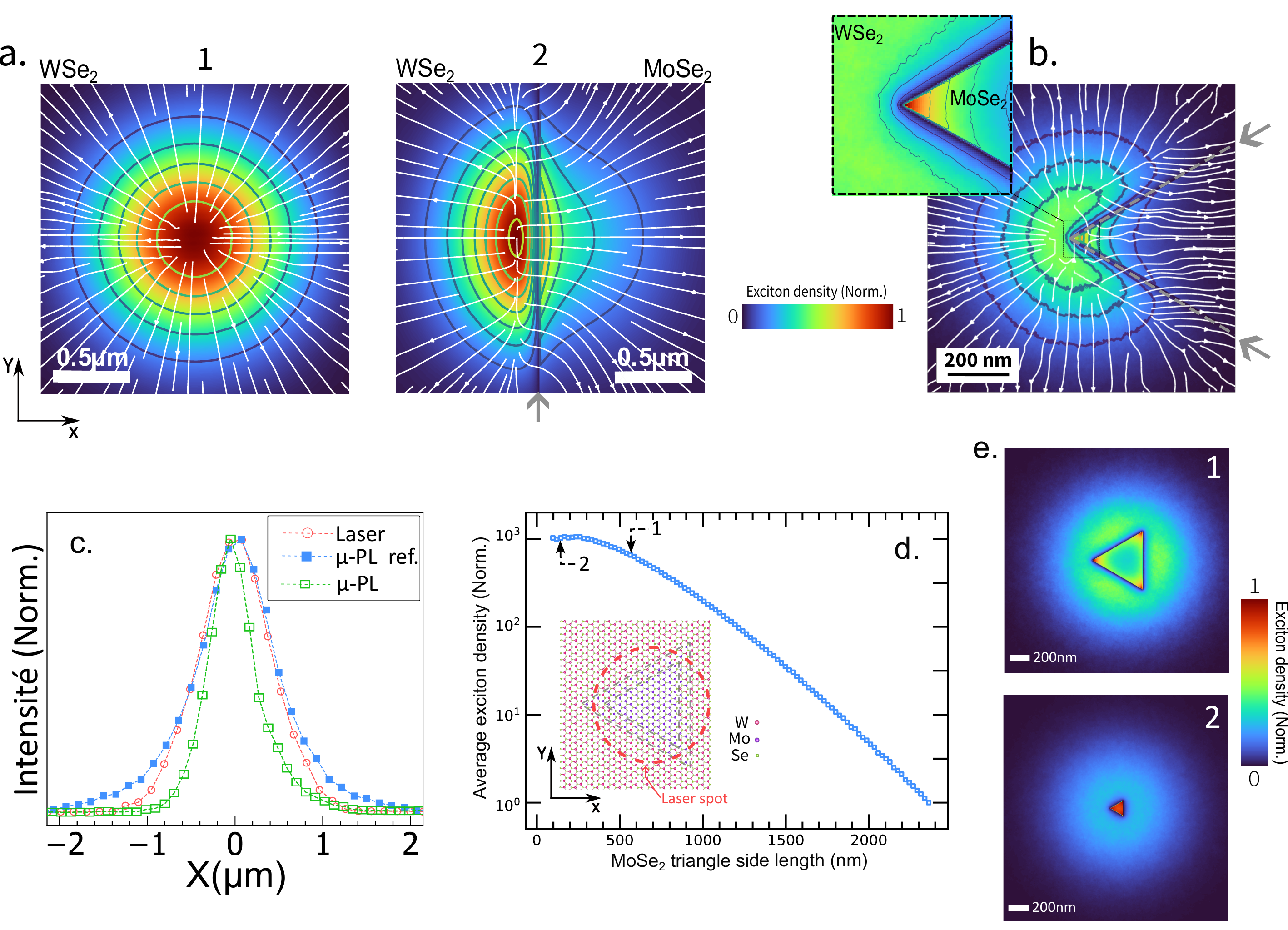}
    \caption{\textbf{Exciton flow at the junction}. a. Exciton density maps calculated for two laser positions: color map 1 inside the WSe$_2$ region far from the interface, and map 2 near the interface. The colored contours represent isodensity lines. White streamlines indicate the orientation of the exciton flux, showing isotropic diffusion far from the junction (map 1) and strongly directed flow near the junction (map 2). The gray arrow marks the junction position. b. Exciton distribution and flow direction for a triangular LH, with the laser source positioned near the apex of the triangle. c. Line profiles extracted from the laser intensity image (red curve) and the spatial PL intensity image (green curve), with the excitation laser centered on a small MoSe$_2$ triangle. For reference, the line profile corresponding to the isotropic case, taken in the MoSe$_2$ monolayer far from any interface, is shown in blue. d. Calculated average exciton density inside the MoSe$_2$ triangle as a function of triangle side length, with the excitation laser centered at the centroid of the MoSe$_2$ triangle, as illustrated in the inset.  e. Simulated exciton density maps for equilateral triangles with sides of (1) 500 nm and (2) 150 nm. For clarity, both color maps are individually normalized. Adapted from Ref. \cite{Lamsaadi2025}}
    \label{ExcitonLensing2}
\end{figure}
Simulation of exciton transport reveals the steady-state distribution of excitons, allowing the prediction of their trajectories. Simulated exciton density maps, combined with current streamlines, illustrate how the junction modifies exciton transport, cf. Figure~\ref{ExcitonLensing2}a. As expected far from the junction, exciton diffusion is isotropic, as evidenced by the radial isodensity colored contours in color map 1. In contrast, in the vicinity of the junction (map 2), the exciton flux becomes strongly redirected and asymmetric. In this forced diffusion regime, a dominant directional exciton current emerges, extending up to 1~µm.\\
Going further, more complex geometries such as triangular MoSe$_2$ tips formed by the intersection of two LHs at a 60$^\circ$ angle have been considered. Simulations of exciton flux reveal that this geometry significantly alters exciton trajectories. When the laser excitation is focused at the triangle apex (cf. Figure~\ref{ExcitonLensing2}b), excitons are funneled along the inclined junctions toward a "focal point" within the MoSe$_2$ region, where an exciton high-density zone emerges. This geometric redirection of exciton paths highlights the role of such LHs as an exciton funnel. Additionally, exciton streamlines become highly collimated and nearly parallel, extending over distances of up to 1.5~µm. Overall, these results demonstrate the potential for guiding exciton flow through precise control of both interface geometry and excitation location.\\
A second key advantage of LHs is their ability to confine neutral excitons within well-defined nanometric regions. A triangular MoSe$_2$ nano-island embedded in a WSe$_2$ monolayer serves as an efficient exciton trap, directing excitons inward and confining them within the triangular domain. As the size of the triangle decreases, from several microns to a few tens of nanometers, the average exciton density within the structure increases by up to three orders of magnitude \cite{Lamsaadi2025}. This behavior has been observed in experiments by imaging the PL emission from sub-wavelength MoSe$_2$ triangles laterally bound to a WSe$_2$ monolayer. Figure~\ref{ExcitonLensing2}c shows the resulting linear profiles of the recorded PL images. 
The PL spot measured in the sub-wavelength triangle (green squares) is significantly smaller than that expected from isotropic diffusion in homogeneous MoSe$_2$ (blue squares), and even smaller than the excitation laser spot (red squares). As observed consistently across all sub-wavelength triangles, the PL profile exhibits a pattern closely matching the Airy disk of a diffraction-limited source, regardless of the physical size of the emitting MoSe$_2$ triangle. Figure~\ref{ExcitonLensing2}d shows the modeled average exciton density in a MoSe$_2$ equilateral triangle as a function of side length, with the laser centered at the triangle’s centroid (see inset). 
As the triangle side length shrinks from 2500 nm to 100 nm, the exciton density increases by three orders of magnitude, illustrating the ability of small MoSe$_2$ regions to attract and trap excitons from surrounding WSe$_2$. In the case of intermediate-sized triangles (cf. color map 1 in Figure~\ref{ExcitonLensing2}e), excitons are focused toward the three corners, forming hot spots due to a lensing-like redirection of exciton fluxes as discussed previously. As the triangle size decreases further (color map 2 in Figure~\ref{ExcitonLensing2}e), these hot spots coalesce into a more uniform exciton density. This transition highlights the ability of sub-wavelength MoSe$_2$ structures to act as highly effective exciton traps and potential building blocks for engineered excitonic systems.
\begin{figure}[b]
    \centering
    \includegraphics[width=1\linewidth]{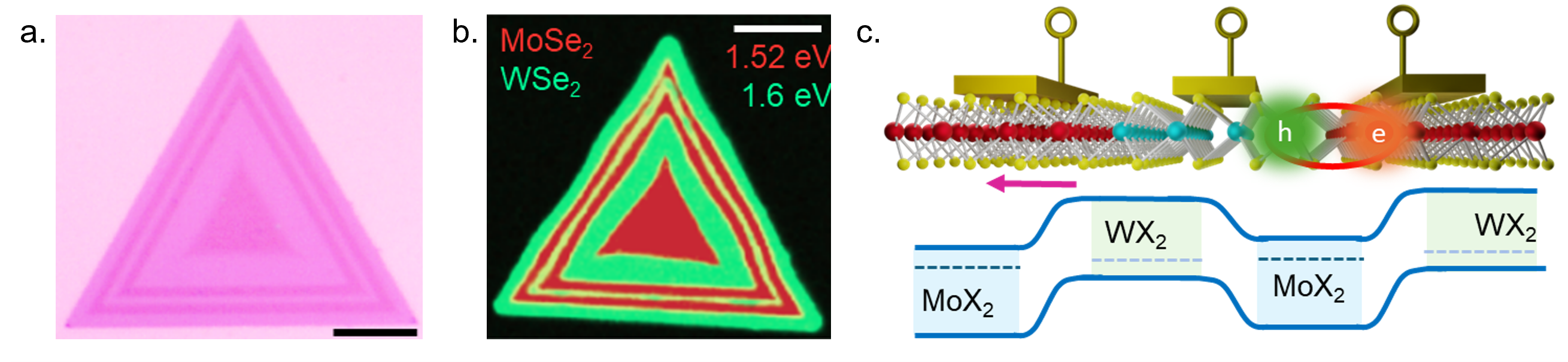}
    \caption{\textbf{Multijunction lateral heterostructures}. a. An optical micrograph of MoSe$_2$–WSe$_2$ LH and b. corresponding PL intensity map, scale bars 10~µm. c. Possible device geometry for controlling exciton transport across multiple interfaces. The bottom shows the electrostatic potential variations. a. and b. are adpated from \cite{Sahoo2018}.}
    \label{Fig-outlook}
\end{figure}
\section{Outlook on lateral heterostructure engineering and exciton physics}

Recent advances in the understanding of lateral junctions in transition metal dichalcogenide monolayers have been propelled by progress in both material synthesis and high-resolution characterization techniques. The quality and spatial extent of these interfaces are critically influenced by growth methods, with chemical vapor deposition playing a pivotal role. Post-growth transfer of CVD-grown heterostructures onto high-quality hexagonal boron nitride substrates has been shown to significantly improve optical and transport properties.\\
Notably, room-temperature tip-enhanced spectroscopy and cryogenic, sub-micrometer resolved optical measurements have enabled deeper insights into exciton behavior and transport across atomically sharp interfaces. These experimental capabilities have set the stage for investigating more complex heterostructures.

In terms of synthesis, extending a single-junction LH into a multi-junction platform—where dissimilar TMDs are sequentially stitched in-plane to form a lateral superlattice—is a central challenge in 2D materials growth cf. Figure~\ref{Fig-outlook}a and b. Recent breakthroughs include the controlled sequential growth of MoX$_2$–WX$_2$ (X = S, Se or both) multi-junction LHs via selective evaporation of metal precursors under distinct carrier gas environments during CVD \cite{Sahoo2019,Sahoo2018}. This method allows precise control over interface width, degree of alloying, and band engineering across 1D interfaces \cite{nugera2022bandgap}, enabling studies such as exciton dissociation in photoconductivity measurements \cite{Berwegerdoi:10.1021/acsnano.0c06745}.

From an optoelectronics perspective, lateral heterostructures offer a rich platform for tunable devices cf. Figure~\ref{Fig-outlook}c. While interlayer excitons in vertical heterostructures have been extensively studied, their in-plane analogs—charge-transfer excitons across lateral junctions—remain relatively unexplored. These CT excitons exhibit an intrinsic electric dipole moment, which enables observation of the Stark shift under an in-plane electric field, a feature absent in monolayer excitons. Additionally, CT excitons are quantized in their center-of-mass motion across the interface while remaining unquantized along it \cite{Rosati2023}. The roles of dark excitons and other quasiparticles in mediating interfacial exciton transport remain open questions that merit further exploration.

Beyond excitons, LHs can host high-density exciton gases and even electron plasmas, as demonstrated in recent work \cite{sousa2025enhanced}. The lateral gradient of alloy composition can function as a confinement potential, enabling the generation of exciton populations with higher densities than those found in pristine monolayers.

The recent observation of 1D dipolar excitonic states points to the potential of LHs as naturally formed 1D excitonic traps with long coherence lengths \cite{Kundu2024,vandoolaeghe25,Rosati2023}. These findings open new directions, including multi-interface exciton coupling, collective oscillation phenomena, and the realization of lateral exciton superlattices that could support room-temperature quantum light sources in planar geometries.

To fully realize this potential, future research could incorporate time-resolved line scans, ultrafast pump-probe spectroscopy, and magneto-optical studies to unravel the dynamics and interactions of interfacial excitonic species. Moreover, angle- and time-resolved photoemission spectroscopy (tr-ARPES) offers a promising route to directly probe band alignment in LHs, essential for understanding charge separation and transport \cite{Madeo2020}. Such measurements would further facilitate the direct tracking of the formation of the charge-transfer excitons analogously to the one of interlayer states in vertical heterostructures \cite{Schmitt22}. The inherent strain due to lattice mismatch between dissimilar TMDs adds another layer of tunability, enabling strain engineering of electronic and optical properties.

Together, these developments establish multi-junction LHs as a frontier for exciton-lattice engineering. The ability to electrically and optically control a wide range of excitonic species opens pathways toward designing artificial 1D exciton lattices with tunable coupling and dispersion, analogous to moiré minibands in vertical systems. These lateral architectures may ultimately enable scalable and planar platforms for quantum optoelectronics and coherent light-matter interactions at room temperature.

\section{References}

\section{Acknowledgements}
HL, VP and JMP acknowledge funding from ANR Ti-P (ANR-21-CE30-0042). AT acknowledges financial support of the DFG through projects TU 149/16-1  (464283495); TU149/21-1 (535253440), CRC NOA 1375, B02 (398816777) and European Union Graphene Flagship project 2D Materials of Future 2DSPIN-TECH (101135853). E.M. and R.R.  acknowledge financial support by the Deutsche Forschungsgemeinschaft  (DFG) via project 512604469. BU, AT, EM, RR and SSh acknowledge financial support by the Deutsche Forschungsgemeinschaft  (DFG) via SPP 2244.

\end{document}